\documentclass{ws-ijmpd}

\newcommand{\beqs}{\begin{equation*}}
\newcommand{\beq}{\begin{equation}}

\newcommand{\eeqs}{\end{equation*}}
\newcommand{\eeq}{\end{equation}}

\newcommand{\beqas}{\begin{eqnarray*}}
\newcommand{\beqa}{\begin{eqnarray}}

\newcommand{\eeqas}{\end{eqnarray*}}
\newcommand{\eeqa}{\end{eqnarray}}

\newcommand{\de}{\delta}
\newcommand{\om}{\omega}

\newcommand{\blist}{\begin{itemize}}

\newcommand{\elist}{\end{itemize}}

\providecommand{\href}[2]{#2}

\DeclareFontFamily{OT1}{rsfs}{}
\DeclareFontShape{OT1}{rsfs}{m}{n}{ <-7> rsfs5 <7-10> rsfs7 <10->rsfs10}{} 
\DeclareMathAlphabet{\mycal}{OT1}{rsfs}{m}{n}

\DeclareMathOperator{\extdm}{d}
\newcommand{\extd}{\extdm \!}

\begin{document}

\markboth{L.~BERGAMIN AND D.~GRUMILLER}
{KILLING HORIZONS KILL HORIZON DEGREES}

\catchline{}{}{}{}{}

\title{KILLING HORIZONS KILL HORIZON DEGREES}

\author{L.~BERGAMIN}

\address{ESA Advanced Concepts Team (EUI-ACT), ESTEC, \\
Keplerlaan 1, NL-2201 AZ Noordwijk, The Netherlands\\
Luzi.Bergamin@esa.int}

\author{D.~GRUMILLER}

\address{Institute for Theoretical Physics, University of Leipzig, \\
Augustusplatz 10-11, D-04109 Leipzig, Germany\\ and \\ 
Center for Theoretical Physics,
Massachusetts Institute of Technology,\\
77 Massachusetts Ave.,
Cambridge, MA  02139\\
grumil@lns.mit.edu}

\maketitle

\begin{history}

\received{13.~September 2006}
\accepted{21.~November 2006}
\comby{D.V.~Ahluwalia-Khalilova}
\end{history}

\begin{abstract}
  Frequently it is argued that the microstates responsible for the
  Be\-ken\-stein--Hawking entropy should arise from some physical
  degrees of freedom located near or on the black hole horizon. In
  this Essay we elucidate that instead entropy may emerge from
  the conversion of physical degrees of freedom, attached to a generic
  boundary, into unobservable gauge degrees of freedom attached to the
  horizon. By constructing the reduced phase space it can be
  demonstrated that such a transmutation indeed takes place for a
  large class of black holes, including Schwarzschild.
\end{abstract}

\keywords{Black hole horizons, Bekenstein-Hawking entropy, edge states, 2D dilaton gravity}

\setcounter{footnote}{0}

\section{Introduction}

Of all the problems quantum gravity is beset with, the most
serious 
is the absence of experimental data --- even if we had the correct
theory how would we know?  Would its Beauty alone reveal its Truth?
Depending on philosophical prejudices sometimes this is answered
affirmatively, but more cautious people invoke Nature as the ultimate
arbitrator. What can be done to satisfy or at least appease the latter
in the absence of experiments?  Fortunately, there are some aspects of
classical and semi-classical gravity that serve as selection criteria.
For instance, a hypothetical theory of quantum gravity failing to
reproduce Einstein's equations, at least in some limit, would be
considered as perverse.  Let us now search for some semi-classical
selection criterion, because reproducing the Einstein equations
certainly is laudable but hardly a milestone.  According to
\cite{Carlip:2006fm} the ``closest thing to experimental data'' we
have at our disposal is the Bekenstein--Hawking law,
\begin{equation}
  \label{eq:essay1}
  S_{BH}=\frac 14 A\,,
\end{equation}
which states that the black hole (BH) entropy $S_{BH}$ equals to a
quarter of the horizon surface area $A$.  This law has been checked by
many independent methods and seems to be so stable against variations
in the underlying assumptions that few people would take a putative
theory of quantum gravity seriously if it failed to reproduce
\eqref{eq:essay1}. In this sense it provides a semi-classical litmus
test for any approach to quantum gravity.  Among the plethora of them
essentially all retrodict the Bekenstein--Hawking law.  As Carlip
pointed out lucidly \cite{Carlip:2006fm}: 
\begin{quote}
  ``In a field in which we do not yet know the answers, the existence
  of competing models may be seen as a sign of health. But the
  existence of competing models that all {\em agree} cries out for a
  deeper explanation.''
\end{quote}

\section{Gauge-to-physics conversion?}

A possible explanation for the universality of the Bekenstein--Hawking
law might stem from near horizon conformal symmetry. This line of
arguments led to numerous ways \cite{Strominger:1997eq}
of recovering \eqref{eq:essay1} by use of the
Cardy formula \cite{Cardy:1986ie}, which yields a direct relation
between entropy and the central charge. Its presence reflects breaking
of the conformal symmetry, so the corresponding
gauge degrees of freedom should be converted, in a Goldstone-like
manner, into physical degrees of freedom 
counted by Cardy's formula.
Thus, although no particular assumption has been made
what the microstates actually are, one is able to establish
\eqref{eq:essay1} in a rather model independent way.

Without a concrete implementation this is just yet another speculative
idea in quantum gravity. Gratifyingly, Carlip has found a way to make
this program work \cite{Carlip:2004mn} in the context of 2D dilaton gravity
\cite{Grumiller:2002nm}.  We cannot avoid to present the 
action \cite{Schaller:1994es},
\begin{equation}
  \label{eq:essay2}
  S_{\rm 2DG}=-\int \left[X^+(\extd-\omega)\wedge e^- + X^-(\extd+\omega)\wedge e^+ + X \extd \om + e^+\wedge e^- \mathcal{V}\right]\,, 
\end{equation}
but we will need very little of it for the purpose of this
Essay.\footnote{For sake of a self-contained text we present some
  of our notation \cite{Grumiller:2002nm}: $e^\pm$ 
  are dyad 1-forms and 
  $\omega$ is essentially the spin-connection.
  The free function in \eqref{eq:essay2}, $\mathcal{V}=X^+X^- U(X) +
  V(X)$, defines the model and depends on the dilaton $X$ and the
  auxiliary fields $X^\pm$. The Schwarzschild BH arises for
  $U=-1/(2X)$ and $V=\rm
  const$; the Witten BH \cite{Witten:1991yr} 
  for $U=-1/X$ and $V\propto X$.}
The basic idea now is to ask questions of the form ``If a BH is present,
what is the probability of this particular physical process?''. 
Thereby, the BH horizon is implemented as an effective boundary of space-time,
which requires specific constraints defining its characteristics.
But as we know since Dirac, constraints change the
structure of phase space.  Carlip could show \cite{Carlip:2004mn} that
due to the presence of ``stretched horizon constraints'' the algebra
of the gauge constraints is changed, as the classical
Virasoro algebra acquires a central charge proportional to the
stretching parameter (i.e., it vanishes in the limit of approaching a
``sharp horizon'').  In this way a noteworthy agreement with
\eqref{eq:essay1} is found and Carlip's attractive
proposal appears to work.

But there is something very puzzling about this result: a ``stretched
horizon'' does not differ essentially from some generic boundary, so
it is not clear in what sense a BH horizon is special. Deriving
\eqref{eq:essay1} certainly is commendable, but deriving
\eqref{eq:essay1} for an {\em arbitrary} surface ``cries out for some
deeper explanation''. In order to even address this issue it is
mandatory to impose sharp horizon constraints rather than stretched ones.


\section{Physics-to-gauge conversion!}

Exactly this task has recently been attacked 
\cite{Bergamin:2005pg} and led to a complementary
picture. Subsequently the surprising result will be recovered and
explained, though we impose a simpler variational principle than
the Gibbons-Hawking prescription applied in \cite{Bergamin:2005pg}.

Variation of the action \eqref{eq:essay2} leads not only to the
equations of motion, but also to the boundary requirements 
\begin{equation}
  \label{eq:essay3}
  X^\mp\de e^\pm_\|=0\,,\qquad X\de\om_\|=0\,, 
\end{equation}
where $\|$ denotes the parallel component along the boundary. This is
seen immediately from \eqref{eq:essay2} by tracing all partial
integrations.  Generically 
\begin{equation}
  \label{eq:essay1001}
  \de e^\pm_\|=\de \om_\|=0
\end{equation}
is the only choice possible, but there exists a crucial exception:
one can prove that a Killing horizon emerges if either $X^+$ or $X^-$
vanish \cite{Grumiller:2002nm}, regardless of the gauge.  Thus, a
consistent set of sharp horizon boundary conditions is given by
\begin{equation}
  \label{eq:essay1002}
  X^-=\de e^-_\|=\de \om_\|=0\,.
\end{equation}
It is remarkable that horizons differ from generic boundaries
already at the level of the variational principle.

The cases \eqref{eq:essay1001} and \eqref{eq:essay1002} are
implemented via boundary constraints besides the primary and secondary
bulk constraints generating gauge symmetries. Quite generically, this
effects the bulk constraints to become second class. However, it is
found that a horizon entails several first class constraints and
consequently gauge symmetries. A study of the latter reveals
diffeomorphisms along the boundary and local Lorentz transformations
to be unrestricted, concurrent with our intuition about null
hyper-surfaces.  This provides another hint that sharp horizons behave
profoundly different from generic boundaries. Still one should be
careful as the relevant contributions to the Poisson brackets between
the constraints have support only at the boundary. Therefore naive
counting arguments cannot be invoked to get the respective numbers of
physical degrees of freedom.

Instead we have to construct the reduced phase space explicitly as
described detailed in \cite{Bergamin:2005pg}.  It is illuminating to
pinpoint the pivotal technical observations: during the construction
most boundary constraints are found to be obsolete as they simply
represent the analytic continuation of the bulk gauge fixing to the
boundary. In this way \emph{all} boundary constraints disappear for a
generic boundary, leaving two free functions which comprise the
(physical) boundary phase space, in accordance with a well-known
result by Kucha{\v{r}} \cite{Kuchar:1994zk}. They can be interpreted
as mass and its conjugate, the ``proper time'' at the boundary.
Indeed, all 2D dilaton gravity models \eqref{eq:essay2} exhibit a
constant of motion \cite{Grumiller:2002nm} directly related to the ADM
mass. 
In order to reduce its boundary value to the core part,
\begin{equation}
  \label{eq:essay4}
  M = \left. X^+ X^- \right|_{\rm boundary}\,,
\end{equation}
the trivial shift and scaling ambiguities inherent to any mass
definition have been fixed by choosing a convenient ground state and
energy scale, respectively.
From \eqref{eq:essay1001} there is no restriction on
$X^+X^-$ at the boundary, so $M$ may fluctuate off-shell.
However, the sharp horizon boundary conditions \eqref{eq:essay1002}
uniquely determine the constraint $M=0$ which is first class and thus
represents a boundary gauge degree of freedom. The ``proper time'' 
is pure gauge as well because of residual Lorentz transformations.
Consequently the physical boundary phase space is empty for horizons.

The conclusion is inevitable: The physical degrees of freedom present
on a generic boundary are converted into gauge degrees of freedom on a
horizon.

\section{Reconciliation and interpretation}

At first glance our results seem to contradict earlier findings, as
 apparent already from the respective section titles. In order to
 reconcile them 
it may be tempting to interpret the disaccord as a manifestation of BH
complementarity \cite{Carlip:2006fm}: for the outside observer the
horizon is not accessible anyhow and thus a stretched horizon could
provide an adequate description consistent with simple thermodynamical
considerations. However, by the same token one could argue that there
should be something wrong with ``sharp'' asymptotic conditions because
no physical observer can get access to the infinite far region.

Evidently a more convincing interpretation is needed. Since the
reduced phase space is non-empty for {\em any} generic boundary it seems
unlikely to us that the corresponding physical degrees of freedom bear
a {\em direct} relation to BH entropy \eqref{eq:essay1}. 
As we have shown this information is lost on a horizon by
transmutation due to enhanced symmetry. Because the very notion of
entropy is built upon (non-accessible) information it is plausible at
a qualitative level why \eqref{eq:essay1} emerges. To obtain it
quantitatively one can still employ the Cardy formula and proceed
e.g.~by analogy to \cite{Carlip:2004mn}, but our results put that
procedure in a quite different perspective.
Though entropy appears to stem from Goldstone modes due to breaking of
conformal symmetry, our analysis demonstrates that 
they actually mimic
physical modes which were there in the first place, but which
have been converted on a horizon into something unobservable, namely
gauge degrees of freedom.
This resonates strikingly with a proposal by 't~Hooft
\cite{'tHooft:2004ek}:
\begin{quote}
  ``A more subtle suggestion is that, although we do have fields
  between the [brick] wall and the horizon, which do carry degrees of
  freedom, these degrees of freedom are not physical. They could
  emerge as a kind of {\em local gauge degrees of freedom},
  undetectable by any observer.''
\end{quote}
It is reassuring that we could establish the validity of this
conjecture for BHs permitting an effective 2D description as in
\eqref{eq:essay2}, including Schwarzschild.  Since also more
complicated scenarios allow such a treatment, at least in the vicinity
of the horizon, the features discussed in this Essay ought to be 
universal for BHs.

To summarize our findings in a single sentence:
\begin{quote} 
  {\em Entropy arises because approaching the black hole horizon does
    not commute with constructing the physical phase space.}
\end{quote}

\section*{Acknowledgment}

We thank cordially Wolfgang Kummer and Dima Vassilevich for collaboration on related topics, in particular on \cite{Bergamin:2005pg}. DG has been supported by project GR-3157/1-1 of the German Research Foundation (DFG) and during the final preparations of this paper also by the Marie Curie Fellowship MC-OIF 021421 of the European Commission under the Sixth EU Framework Programme for Research and Technological Development (FP6).



\providecommand{\href}[2]{#2}\begingroup\raggedright\endgroup

\end{document}